\documentclass[floatfix,amsmath,amssymb,aps,twocolumn]{revtex4}
\usepackage{times}
\usepackage[dvips]{epsfig}
\begin{document}
\title{Why is Random Close Packing Reproducible?}

\author
{Randall D. Kamien}
\author{Andrea J. Liu}
\affiliation{Department of Physics and Astronomy, University of Pennsylvania,
Philadelphia, PA, 19104, USA}

\date{\today}

\begin{abstract}
We link the thermodynamics of colloidal suspensions to
the statistics of regular and random packings.  Random close packing has defied a rigorous definition yet, in three dimensions, there is near universal agreement on the volume fraction
at which it occurs.  We conjecture that the common value of $\phi_{\rm rcp}\approx 0.64$ arises from
a divergence in the rate at which accessible states disappear.  We relate this rate to the equation
of state for a hard sphere fluid on a metastable, non-crystalline branch.
\end{abstract}

\maketitle 
When Bernal turned to sphere packings as a route towards understanding liquids \cite{0a}, he recognized the importance of the {\it random close-packing} density, $\phi_{\rm rcp}$.  This density was defined operationally to be the fraction of occupied volume in the densest disordered packing of hard spheres.   Early measurements of $\phi_{\rm rcp}$ by Scott~\cite{0b} and Bernal and Mason~\cite{0c} involved pouring, shaking and kneading ball bearings in flasks and balloons.   Since then, these measurements have been reproduced by countless experiments and numerical algorithms, which find a common value of $\phi_{\rm rcp}\approx 0.64$ in three dimensions~\cite{aste}.  There are therefore many distinct but numerically-consistent definitions of $\phi_{\rm rcp}$ that depend on procedure.  At the time of Scott's experiments, Bernal and Mason wrote, ``The figure for the occupied volume of random close packing--0.64--must be mathematically determinable, although so far as we know undetermined.''~\cite{0c}.  Nearly fifty years later, however, a mathematically-rigorous definition of random close-packing still remains elusive.  Why should an ill-defined state have a reproducibly accepted value for its volume fraction? 

Two approaches towards a mathematically-rigorous definition have been suggested recently.  Torquato, {\sl et al.} \cite{9} pointed out that there is an inherent discrepancy in the concept of a ``random close-packed state,'' because one can always obtain a more  closely-packed state (a denser state) by introducing order into the system.  This must be true because $\phi_{\rm rcp}$ is lower than the maximum close-packed density~\cite{hales}, $\phi_{\rm FCC}=\pi/3\sqrt{2} \approx 0.74$, corresponding to the close-packed FCC crystal.  Torquato, {\sl et al.} \cite{9} proposed an alternate way of thinking about random close-packing, in terms of  a ``maximally-random jammed'' (MRJ) state at $\phi_{\rm rcp}$.   ``Maximally random'' states or configurations are those with minimal values of typical order parameters, such as bond-orientational order or crystalline order.  ``Jammed'' states have the property that any particle or set of particles cannot be translated with respect to any of the rest of the particles in the system without introducing overlaps.  Kansal, {\sl et al.} \cite{9a} showed that several different order parameters yielded consistent estimates for the packing fraction of the maximally random jammed state, $\phi_{\rm MRJ} \approx 0.63$, in good agreement with $\phi_{\rm rcp}$.  This is very encouraging.  Nonetheless, there is some uncertainty in this approach associated with the order parameter; one cannot calculate all possible order parameters, or determine that all order parameters yield the same MRJ packing fraction.   
Moreover,  there may be a specific type of order associated with the MRJ state~\cite{7a,9b,7c,Anikeenko}, which would then be maximal rather than minimal in the MRJ state, though these special metrics are likely to be of a different nature from those studied in \cite{9}.

A second approach by O'Hern, {\sl et al.} is based on the energy landscape of systems of soft spheres. Specifically, Refs.~\cite{7,7a} considered potentials of the form
 \begin{equation} 
 \label{potentialdef} 
 V(r) = \left\{ \begin{array}{cc} 
 \epsilon (1-r/\sigma)^\alpha/\alpha & {\rm for}~r < \sigma \\ 
 0 & {\rm for}~r \ge \sigma 
 \end{array} 
 \right. 
 \end{equation} 
Here, $\epsilon$ is the characteristic energy of interaction and $\sigma$ is the particle diameter.  For $\alpha=3/2$, $\alpha=2$ and $\alpha=5/2$, O'Hern, {\sl et al.} studied the fraction of ideal gas states $f_0(\phi)$ that belong to basins of attraction corresponding to zero energy states (or equivalently, allowed hard sphere states).  It was found that $-d f_0(\phi)/d\phi$, {\sl i.e.} the rate at which the fraction of ideal gas states belonging to basins of attraction of hard sphere states shrinks with increasing density, appears to develop a delta-function peak at $\phi_{\rm rcp}$ in the infinite system size limit~\cite{7,7a}.  Thus, they suggest that in the thermodynamic limit, the vast majority of ideal gas states belong to basins of attraction of hard sphere states that jam at $\phi_{\rm rcp}$.   Here, a state is ``jammed'' if there are no zero frequency vibrational modes except for those due to floaters (particles with no overlapping neighbors) or overall translation and rotation of the system~\cite{remark}.    The uncertainty in this approach is associated with how the energy landscape is explored~\cite{9c,7d}; different algorithms may yield different final energy minima for a given ideal gas state.

In this paper, we exploit a relation that has been used to calculate the pressure of a system to explore the connection between the pressure -- a thermodynamic quantity that can be measured for colloidal suspensions -- and the behavior of hard sphere packings.  In particular, we show that the fractional rate at which allowed states disappear with increasing volume fraction is proportional to the pressure.   We also show that free volume theory provides a reasonable fit to the equation of state of the hard sphere liquid, with a single fit parameter $\phi_{\rm max}$, corresponding to the density at which the pressure diverges.  We find that $\phi_{\rm max} = 0.640 \pm .006$, in excellent agreement with $\phi_{\rm rcp}$.   We conclude by discussing these results in the context of previous work and making some conjectures regarding the origin of the reproducibility of $\phi_{\rm rcp}$.

We first derive a useful relation between the pressure and the rate at which allowed states disappear.  This relation has been used to calculate the pressure from a Monte Carlo run for hard particles by calculating the minimum density change needed to introduce overlaps between particles~\cite{0.1}.  Here, we use the relation to gain insight into the packings of hard spheres.  We consider the number of available states as a function of volume fraction $\phi=\rho v$, where $\rho$ is the number density of spheres and $v$ is their volume.  We consider the probability of finding an allowed hard sphere state at $\rho$, 
\begin{equation}
{\cal R}(\rho)=Z_{\rm HS}(\rho)/Z_{\rm IG}(\rho) ,\label{Rdef}
\end{equation}
where $Z_{\rm HS}$ and $Z_{\rm IG}$ are the numbers of allowed configurations (or equivalently, the canonical partition functions) for hard spheres and for an ideal gas of point particles, respectively.   We now consider the effect of increasing the volume fraction of a particular state at density $\rho$.  The probability of {\sl not} finding a new state at $\rho+\delta \rho$, given this state at $\rho$, is $J(\rho)\delta \rho$, where
$J(\rho)=-\frac{1}{{\cal R}} \frac{\partial {\cal R}}{\partial \rho}$.  Using the thermodynamic definition of pressure, $p \equiv T \partial \ln Z/\partial V$, we find 
$J(\rho) = \frac{V}{\rho T}\left[p_{\rm HS}(\rho)-p_{\rm IG}(\rho)\right],$
where $p_{\rm HS}$ and $p_{\rm IG}$ are the pressures of the hard sphere and ideal gas systems, respectively, $V$ is the volume and $T$ is the temperature (the Boltzmann constant is unity).  Equivalently,
\begin{equation}
J(\phi)=\frac{V}{\phi T} \left[p_{\rm HS}(\phi)-\phi T/v \right]. \label{Jdef}
\end{equation}

In Fig.~1(a) we show the experimentally-measured equation of state for a hard sphere system (circles) \cite{1}.  The pressure increases with $\phi$ until $\phi_{\rm X} \approx 0.49$, at which point fluid and crystal begin to coexist.  At $\phi\approx 0.54$ the system is completely crystalline and $p_{\rm HS}(\phi)$ eventually diverges at $\phi_{\rm FCC}=\pi/3\sqrt{2} \approx 0.74$, the packing fraction of the close-packed FCC lattice.  Eq. (\ref{Jdef}) implies that $J(\phi)$ also diverges at $\phi_{\rm FCC}$, signaling an inability to construct any higher density states, {\sl i.e.} close packing.   Thus, if all the states are counted in $Z_{\rm HS}$, $J$ does not diverge until $\phi_{FCC}$.   

\begin{figure}
\centering
\includegraphics[width=\columnwidth]{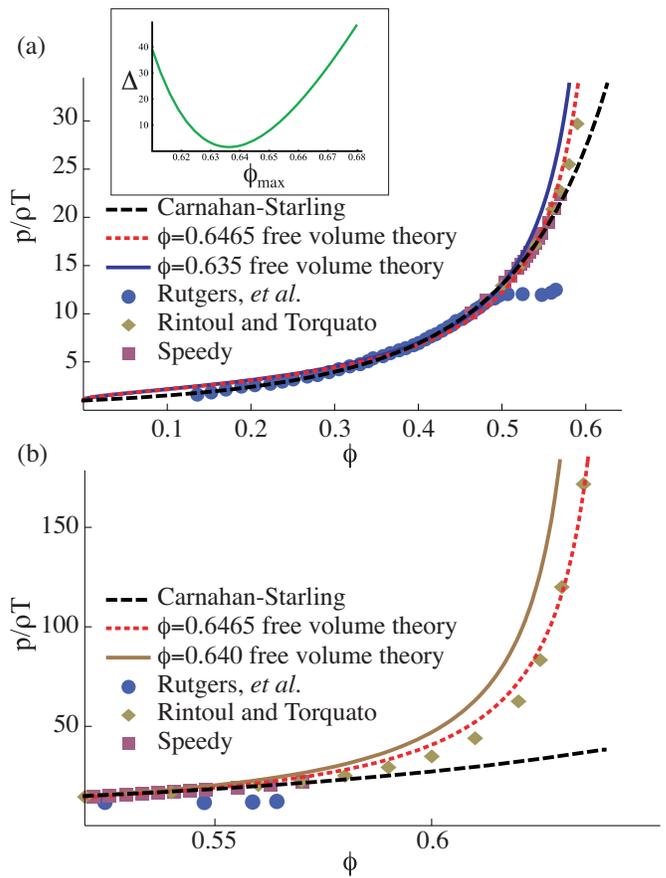}
\caption{(Color Online) Measured equation of state $p$ (in units of $\rho T$) versus volume fraction $\phi$ for a hard sphere fluid from experiments by Rutgers, {\sl et al.} \protect{\cite{1}} (circles) and simulations by Rintoul and Torquato \protect{\cite{2}} (diamonds) and Speedy \protect{\cite{2.1}} (squares).  The dashed curve is the Carnahan-Starling equation of state.  In a) we show two fits to free volume theory, corresponding to $\phi_{\rm max}=0.635$ (solid), which is the best fit to the experimental data for the equilibrium liquid branch, and $\phi_{\rm max}=0.6465$ (dotted), which is the best fit to the numerical data for the metastable branch.  In b) we show free volume theory for $\phi_{\rm max}=0.6465$ (dotted) and $\phi_{\rm max}=0.640$ (solid). Part (a) of the figure shows the quality of the fit to the equilibrium liquid branch while part (b) shows the quality of the fit to the metastable branch.  Inset to (a): the variation in the mean-square error $\Delta$ between free volume theory and the experimental data as a function of $\phi_{\rm max}$.  }
\end{figure}

There is also a metastable branch of the equation of state that is reproducibly measurable numerically (see Fig.~1) \cite{2,2.1}\ and potentially measurable experimentally in rapidly sedimented colloidal suspensions \cite{4}.  This branch is used to study the colloidal glass transition \cite{3}\ and dynamics in granular media \cite{5}.  Along this branch, the pressure apparently diverges\cite{2,2.1} at $\phi_{\rm rcp}$ as
$p_{\rm HS}(\phi) \sim (\phi_{\rm rcp}-\phi)^{-1}$.  Such a divergence is predicted on the basis of polytope theory, or free volume theory, for classical hard spheres~\cite{stillinger,9b}.  According to Eq. (\ref{Jdef}), this would lead to a divergence in $J$ there.  To construct an analytical approximation to a metastable branch that continues smoothly from $\phi_{\rm X}$ to a divergence at some $\phi_{\rm max}$, we therefore turn to free volume theory.    
To calculate the free energy within free volume theory, we must construct the Voronoi tesselation for any particular packing \cite{6}.  For a given packing, each sphere is allowed to independently explore the free volume of its cell.  The free energy involves a sum over all possible tesselations and requires knowledge of the distribution of free volumes \cite{5.5}.  However, the change of free energy with $\phi$, or the pressure, should not depend on the tesselation sufficiently close to $\phi_{\rm max}$.  If $\phi$ is decreased infinitesimally near $\phi_{\rm max}$, the increase in free volume of each cell for an isotropic, affine rarefaction of the packing scales as $\left[(\phi_{\rm max}/\phi)^{1/3}-1\right]^3$ as long as the shape of each free volume region remains fixed.   Thus, near $\phi_{\rm max}$, the pressure depends on only one free parameter, $\phi_{\rm max}$:
\begin{equation}\label{RCPeos}
p_{\rm FV}=-\frac{T\phi^2}{v}\frac{d}{d\phi}\ln\{[(\phi_{\rm max}/\phi)^{1/D} -1]^D\}
\end{equation}
where $D$ is the dimension of space \cite{6}.  Note that as the volume fraction further decreases, the geometry of the allowed volume will change and, though the scaling factor remains the same, the prefactor will change, leading to additional corrections.  Recent arguments~\cite{HR} based on statistical geometry suggest that Eq. (\ref{RCPeos}) should be at least approximately correct for the metastable branch.  

Fig.~1(a) shows that free volume theory provides a very reasonable fit to the equilibrium liquid branch of the data of Ref.~\cite{1} just below $\phi_{\rm X}$, with $\phi_{\rm max}\approx 0.64$ in 3D, in excellent agreement with measured RCP values.  The inset shows that $\Delta$, the mean-squared error, is a strong function of $\phi_{\rm max}$.  We find a comparable value of $\phi_{\rm max} \approx 0.636$ when we fit the range $\phi\in[0.3,0.5]$ to the first ten virial coefficients \cite{Clisby}.  Unsurprisingly, we also find a similar value ($\phi_{\rm max}\approx 0.635$) when we fit to the Carnahan-Starling approximate equation of state \cite{CS}.  The latter function is known to describe experimental or numerical measurements of the liquid branch of the equation of state to within measurement error~\cite{Piazza,1,Phan,Gruhn}.  It is worth noting that the fitted values of $\phi_{\rm max}$ are insensitive to the region over which we fit the data.  Using the entire equilibrium liquid branch $\phi\in[0,0.5]$ only changes $\phi_{\rm max}$ in the last decimal place when fitting to the experimental data, the virial expansion, or the Carnahan-Starling approximation (for example, we obtain $\phi_{\rm max}\approx 0.637$ when fitting to the Carnahan-Starling equation over the whole range).    Aste and Coniglio~\cite{6a}\ have used local random configurations of hard spheres to estimate the pressure in free volume theory and there, too, find an independent branch of the equation of state diverging at $\phi_{\rm max}\approx 0.65$.

We have also fit free volume theory to numerical measurements of the metastable branch~\cite{2,2.1}.  Here we find that the best fit corresponds to $\phi_{\rm max} \approx 0.6465$--less than 2\% higher than the result for the fit to the equilibrium liquid branch.   In Fig.~1(a), we plot Eq. (4) for both $\phi_{\rm max}=0.635$ (solid) and $\phi_{\rm max}=0.646$ (dotted); these curves are nearly indistinguishable in terms of their ability to fit the equilibrium portion of the curve.  In Fig.~1(b), we plot the same data and fits over a narrower range of $\phi$ and larger range of pressure to show the quality of the fit to the metastable branch.  Note that our one-parameter fit reproduces not only the shape of the divergence but the overall amplitude of the pressure reasonably well.  The final value $\phi_{\rm max}=0.640 \pm 0.006$ represents the range of fit values that we obtain, though the metastable branch strongly favors the high
end of this range, as seen in Fig.~1(b).

We now note that the many different algorithms and preparation methods that yield $\phi_{\rm rcp} \approx 0.64$, have one point in common:  they all contrive to avoid crystallization and should thus correspond to a metastable branch of the pressure.  For two very different approaches, conjugate gradient minimization \cite{7,7a}\ and the Lubachevsky-Stillinger algorithm \cite{LS,9b}, there is good evidence that the accessible states follow the branch estimated in Fig.~1.  Both procedures yield a pair correlation function $g(r)$ that diverges at contact as $\vert\phi_{\rm rcp}-\phi\vert^{-1}$, implying a likewise diverging pressure.  These approaches both observe a catastrophic loss of states, {\sl i.e.} they jam, at $\phi_{\rm rcp}$, consistent with Eq. (1).  

These results suggest that  $\phi_{\rm rcp}$ corresponds to a divergent endpoint of a metastable branch of the equation of state for hard spheres.  This provides another way of thinking about why random close-packing has not been a well-defined concept \cite{9}.  It is well understood that metastable branches are somewhat arbitrary.  To obtain metastable branches, one must introduce additional constraints that effectively exclude certain states (such as crystalline states) from the partition function \cite{9.6}.  Different choices of constraints lead to different metastable branches.  Alternatively, one can think in terms of the Andreev-Fisher droplet model \cite{9.2,9.4}, or the instanton approach to first-order transitions \cite{9.5}, which predict an essential singularity at the onset of crystallization, $\phi_X$.  This essential singularity precludes analytic continuation of the pressure beyond $\phi_X$; physically, droplets of the nucleating crystalline phase prevent the clean definition of the metastable branch. 

The possibility of multiple metastable branches leads to the possibility of multiple divergent endpoints.  The existence of multiple divergent endpoints is strongly suggested by results of Torquato, {\sl et al.} \cite{9} using the Lubachevsky-Stillinger algorithm~\cite{LS}.  In this algorithm, one starts with an equilibrium liquid configuration at low volume fraction and compresses by increasing the diameters of all the particles at some rate $\Gamma$.  The system jams at some packing fraction $\phi_f(\Gamma)$, which approaches $\phi_{\rm rcp}$ from above in the limit $\Gamma \rightarrow \infty$.  This suggests that one can find metastable branches that end at any value of $\phi$ between $\phi_{\rm rcp}$ and $\phi_{\rm FCC} \approx 0.74$.

Torquato, {\sl et al.} \cite{9} have suggested that the state at $\phi_{\rm rcp}$ is special in the sense that it corresponds to a ``maximally random jammed'' state~ \cite{9}.  It seems reasonable that this concept should be generalizable to an entire ``maximally-random jammed'' metastable branch of the pressure that ends at $\phi_{\rm rcp}$.   This is the philosophy underlying calculations of 
Rintoul and Torquato \cite{2}, who followed a metastable branch of the pressure by discarding all states with appreciable values of the bond orientational order parameter.  In that case, they found a diverging pressure at $\phi_{\rm rcp}$.

Another possibility that is not inconsistent with the scenario of Torquato, {\sl et al.} is that $\phi_{\rm rcp}$ might be better defined than the metastable branch that it terminates.  It might correspond to a true singularity of the free energy that is inaccessible in equilibrium due to the essential singularity at $\phi_X$, the onset of crystallization.   Exact analyses of one-dimensional models show that it is possible for a system to have a metastable branch that is not well-defined but that ends in a divergent endpoint that is a true singularity~\cite{10}.  

We therefore conjecture that $\phi_{\rm rcp}$ represents a special well-defined divergent endpoint of a set of metastable branches of the pressure.  Any procedure that samples a nonzero fraction of states belonging to metastable branches with this endpoint will yield a divergent pressure, and therefore a divergent rate at which states disappear, at $\phi_{\rm rcp}$.   If this conjecture is correct, it would explain why so many different procedures, all sampling states somewhat differently, yield the same value of $\phi_{\rm rcp}$.   At the same time, it is clear that other procedures might yield different jamming densities by avoiding states that belong to metastable branches with endpoints at $\phi_{\rm rcp}$.

Based on results for hard spheres obtained from replica theory, Parisi and Zamponi~\cite{11,11a} have suggested a similar but more elaborate scenario.  In this picture, $\phi_{\rm rcp}$ is the divergent endpoint of a metastable branch, but there is also an another point on the branch, $\phi_{\rm g}<\phi_{\rm rcp}$, which marks a thermodynamic glass transition.  Above $\phi_{\rm g}$, the configurational entropy vanishes; the system must remain in the lowest free energy states and the relaxation time is infinite.  This scenario is not inconsistent with our conjecture.  

A singularity in the rate of change of number of states does not
necessarily imply that most initial states will flow to $\phi_{\rm rcp}$.  It is possible that most states might have their jamming thresholds at values of $\phi$ below $\phi_{\rm rcp}$, leaving only a few that terminate at $\phi_{\rm rcp}$.  Numerical results suggest that the opposite might be true.  Indeed, they suggest that $\phi_{\rm rcp}$ may not only mark a well-defined divergent endpoint of a set of metastable equations of state, but that an even stronger condition might hold: 
 the distribution of jamming thresholds may actually have a divergent maximum at $\phi_{\rm rcp}$ in the thermodynamic limit.   This conjecture is motivated by results of O'Hern, {\sl et al.} \cite{7,7a}, which suggest that for several soft repulsive potentials, the overwhelming number of ideal gas states belong to basins of attraction of hard sphere states that have their jamming thresholds at $\phi_{\rm rcp}$.  Note that the probability of belonging to a basin of attraction of a state with a jamming threshold at $\phi$ depends on both the size distribution of basins of attraction and the distribution of jamming thresholds.  For small systems, Xu, {\sl et al.} \cite{XBO} have separated the these two distributions by explicitly enumerating the jamming thresholds; their results suggest that the distribution of jamming thresholds is maximal near $\phi_{\rm rcp}$, consistent with this conjecture.   While it is unlikely that the maxima of the distributions of jamming thresholds and of jamming thresholds weighted by their basin of attractions and the divergent endpoint of the metastable branch all coincide at exactly the same density, it is possible that they agree with $\phi=0.64$ to within 1-2\%, so that any one of these could constitute a reasonable definition of $\phi_{\rm rcp}$ \cite{Cates}.  
 
The quantity ${\cal R}(\phi)$ defined in Eq. (\ref{Rdef}), the ratio of hard sphere states to the total number of ideal gas states at a given $\phi$, includes not only hard sphere states at their jamming thresholds at $\phi$ but also hard sphere states below their jamming thresholds.  In this paper, we have argued that the form of the equilibrium equation of state suggests that at $\phi_{\rm rcp}$, the latter states completely dwarf the number of hard sphere states that are at their jamming thresholds, hiding the divergence in the pressure.   It is only when these more ordered states are excluded by restricting the system to a metastable branch of the pressure that a signature of $\phi_{\rm rcp}$ appears.  This reasoning suggests that a clean definition of $\phi_{\rm rcp}$ will rely not only on the distribution of jamming thresholds, but the {\it number} of allowed hard sphere states at their jamming thresholds at $\phi_{\rm rcp}$ as a function of system size~\cite{DF}.   

We thank M.E. Cates, P. M. Chaikin, D. J. Durian, M. E. Fisher, D. Frenkel, J. Kurchan, T. C. Lubensky, C. Modes,  S. R. Nagel,  H. Reiss and S. Torquato for discussions.  This work was supported by NSF Grants DMR05-47230 (RDK), DMR-0605044 (AJL), and DOE Grant DE-FG02-05ER46199 (AJL).  We thank the Aspen Center for Physics for their hospitality while this work was being done.

\end{document}